\documentclass[oldversion,preprint]{aa}  
\usepackage{graphicx}
\usepackage{txfonts}
\begin{document}
   \title{Spurious source generation in mapping from noisy 
          phase-self-calibrated data}


   \author{I. Mart\'{\i}-Vidal
          \inst{1}
          \and
          J. M. Marcaide\inst{1}
          }

   \offprints{I. Mart\'{\i}-Vidal}

   \institute{Dpt. Astronomia i Astrof\'{\i}sica, Universitat de Val\`encia,
              C/Dr. Moliner 50, 46100 Burjassot (Valencia), SPAIN\\
              \email{I.Marti-Vidal@uv.es}
             }

   \date{Accepted on 12 December 2007}
 
  \abstract
{Phase self-calibration (or {\em selfcal}) is an algorithm often used in 
the calibration of interferometric observations in astronomy. 
Although a powerful tool, this algorithm presents strong limitations 
when applied to data with a low signal-to-noise ratio. We analyze the 
artifacts that the phase selfcal algorithm produces when applied to 
extremely noisy data. We show how the phase selfcal may generate a 
spurious source in the sky from a distribution of completely random 
visibilities. This spurious source is indistinguishable from a real one. 
We numerically and analytically compute the relationship between 
the maximum spurious flux density generated by selfcal from noise 
and the particulars of the interferometric observations. Finally, we 
present two simple tests that can be applied to interferometric data for 
checking whether a source detection is real or whether the source is an 
artifact of the phase self-calibration algorithm.}


\authorrunning{I.~Mart\'{\i}-Vidal et al.}
\titlerunning{Spurious source generation from noisy self-calibrated data.}

   \maketitle
%

\section{Introduction}

Phase self-calibration (or {\em selfcal}) is an algorithm often 
used in the calibration of radio astronomical data. It was introduced by 
Readhead \& Wilkinson (\cite{Readhead1978}) and Cotton (\cite{Cotton1979}), 
and it has been essential for the success of Very Long Baseline 
Interferometry (VLBI) imaging. Also, the antenna-based calibrations obtained 
from the {\em Global Fringe Fitting} algorithm (Schwab \& Cotton 
\cite{Schwab1983}) are equivalent to a phase self-calibration. 
The phase selfcal will also be an algorithm widely used with future 
interferometric instruments, such as the Atacama Large Millimeter Array 
(ALMA) or the Square Kilometre Array (SKA), now under construction or 
planned. Optical interferometric observations (like those in the Very Large 
Telescope Interferometry, VLTI) will also eventually benefit from some form 
of selfcal, although closure phases and amplitudes are measured in 
optical interferometry in a very different way than in radio. Thus, 
the statistical analysis presented here may need some substantial changes 
to rigorously describe the probability of false detections by optical 
interferometers.

Given that part of the interferometric observations obtained from all those 
instruments may come from very faint sources, it is important to take into 
account the undesired and uncontrollable effects that the instrumentation 
and/or the calibration and analysis algorithms applied to the data could 
introduce in the interferometric observations. A deep study of all our 
analysis tools and their effects on noisy data is essential for discerning 
the reliability of detections of very faint sources.
Some discoveries made by pushing the interferometric instruments to their 
sensitivity limits could turn out to be the result of artifacts produced by 
the analysis tools. 

The main limitations of the phase self-calibration algorithm have been 
analyzed in many publications (e.g., Linfield \cite{Linfield1986}, 
Wilkinson et al. \cite{Wilkinson1988}). It is well known that an unwise 
use of selfcal can lead to imperfect images, even to the generation of 
spurious source components, elimination of real components, and deformation 
of the structure of extended sources. In this paper, we focus on the 
effects that phase self-calibration produces when applied to pure noise. We 
show that selfcal can generate a spurious source from pure noise, 
with a relatively high flux density compared to the rms of the visibility 
amplitudes. We analytically and numerically study how the recoverable 
flux density of such a spurious source depends on the details of the 
observations (the sensitivity of the interferometer, the number of antennas, 
and the averaging time of the selfcal solutions). Finally, we study the 
effects of phase self-calibration applied to the visibilities resulting from 
observations of real faint sources, instead of pure noise. We 
present two simple tests that can be applied to real data in order to check 
whether a given faint source is real or not, and apply these tests to 
real data, corresponding to VLBI observations of the radio supernova 
SN\,2004et (Mart\'i-Vidal et al. \cite{MartiVidal2007}).

\section{Basics of phase self-calibration}

The basics of phase self-calibration can be found in many publications 
(e.g., Readhead \& Wilkinson \cite{Readhead1978}, Schwab \cite{Schwab1980}).
Here, we explain the essence of this algorithm in a few lines. Let us 
suppose that we have made an interferometric observation using a set of $N$ 
antennas. We obtain one visibility, $V_{ij}$, for each baseline, that is, 
for each pair of antennas $(i,j)$. Let us call $\phi_{ij}$ the phase of the 
visibility $V_{ij}$. Any atmospheric or instrumental effect on the optical 
path of the signals that arrived to antennas $i$ and $j$ will affect the phase 
$\phi_{ij}$ in the way:

\begin{equation}
\phi_{ij} = \phi^{str}_{ij} + \phi^{atm}_{i} - \phi^{atm}_{j} ,
\end{equation}

\noindent where $\phi^{atm}_{l}$ represents all the undesired (i.e., 
atmospheric and instrumental) contributions to the optical path of the signal 
received by the antenna $l$ and $\phi^{str}_{ij}$ the contribution to the 
phase that comes purely from the structure of the observed source, 
that is, the so called {\em source structure phase}. It can be easily shown 
that the quantities known as {\em closure phases}, and defined as 
(Jennison \cite{Jennison1958}, Rogers et al. \cite{Rogers1974}):

\begin{equation}
C_{ijk} = \phi_{ij} + \phi_{ik} - \phi_{jk}
\label{ClosPhase}
\end{equation}

\noindent are independent of $\phi^{atm}_{l}$. That is, the closure phases 
$C_{ijk}$ are only defined by the structure of the observed source. Thus, 
they are independent of any atmospheric or instrumental contribution that 
may affect the signals received by the antennas of the interferometer.
The phase self-calibration algorithm takes advantage of the closure phases 
to estimate the undesired antenna-dependent contributions 
$\phi^{atm}_{l}$. In short, the phase selfcal finds which set of 
antenna-dependent quantities $\phi^{gain}_{l}$ (called {\em phase gains}) 
generate the set of phases:

\begin{equation}
\phi^{self}_{ij} = \phi_{ij} - \phi^{gain}_{i} + \phi^{gain}_{j} ,
\end{equation}

\noindent where $\phi^{self}_{ij}$ are the phases that better represent the 
source structure given by the closure phases $C_{ijk}$. In the ideal case, 
$\phi^{gain}_{l} = \phi^{atm}_{l}$ and $\phi^{self}_{ij} = \phi^{str}_{ij}$. 
The process from which the values $\phi^{gain}_{l}$ (for $l = i,j$) are 
obtained is called {\em hybrid mapping}, and its 
explanation can be found in many publications (e.g., Cornwell \& Wilkinson 
\cite{Cornwell1981}). Here, suffice to say that the phase gains of the 
antennas are obtained from a least-square fit of the raw visibilities 
to the source model obtained from the mapping.

The hybrid mapping is an iterative process from which the structure of the 
source model is refined step by step. Often, the model used in the first
iteration of hybrid mapping is a point source located at the center of 
the map (obviously, the flux density of this point source will not affect 
the phase calibration). The successive steps of hybrid mapping and selfcal 
correct this point source model until the structure that better 
represents all the closure phases is obtained.

\section{Probability distribution of the visibilities due to pure noise}

When an interferometer observes a given source with a flux density well 
below the sensitivity limit of its baselines (that is, when the 
interferometric data contain only noise), both the real and imaginary 
parts of the resulting visibilities follow Gaussian distributions, centered 
at the origin. The amplitudes and phases of the visibilities follow 
distributions different from Gaussian. 
It can be shown that, for each baseline, the probability distribution of 
the phases is uniform between $-\pi$ and $\pi$, and that instead the 
distribution of the amplitudes is given by:

\begin{equation}
g(A) = \frac{A}{\sigma_{ij}^2} 
\exp{\left (-\frac{A^2}{2 \sigma_{ij}^{2}} \right )}
\label{ProbDistri1}
\end{equation}

\noindent where $g(A)$ is the probability density of the amplitudes and 
$\sigma_{ij}$ is the width of the Gaussian distributions of the real and 
imaginary parts of the visibilities of the baseline $(i,j)$. The width 
$\sigma_{ij}$ is related to the thermal noise of the baseline $(i,j)$. 
We assume, for simplicity, that all the baselines of the 
interferometer have the same value of $\sigma$. It can be shown that 
the rms of the visibility amplitudes of a pure noise signal is 
$\rho = \sqrt{2} \, \sigma$ and that the mean amplitude, $<A>$, is 
$\sqrt{\pi/2} \, \sigma$, different from zero.

\section{Probability of generating a spurious source from pure noise}

Given that the real part of a visibility with phase in the range 
$(-\pi/2, \pi/2)$ is positive, all the visibilities with phases in that 
range bring a positive mean flux density to the map. We call 
{\em phase close to zero} to a phase in the range $(-\pi/2, \pi/2)$ and 
{\em phase distant from zero} to a phase outside that range.

The distribution of closure phases is uniform between $-\pi$ and 
$\pi$, as it is also the case for the distribution of phases. This means 
that there is a subset of closure phases that by chance are
close to zero, being compatible with a point source. However, there are
also closure phases distant from zero, which are totally 
incompatible with a compact source. If self-calibration is not applied to 
the data, then the uniform distribution of phases (and closure phases) will 
result in a noisy map with no source defined in it. But if a single 
iteration of phase self-calibration is applied, there is a selection 
process of the closure phases in the calibration, which may generate a 
spurious source with a flux density comparable to the rms of the amplitudes
(which can be much higher than the rms of the image), as we 
show below.

For each scan, the effects of the least-square fit described in Sect. 
2 can be understood in the following way: selfcal searches for the 
visibilities corresponding to the antennas most commonly appearing in the 
closure phases that are close to zero (which usually correspond to phases 
that can be modelled with a point source). Then, selfcal minimizes 
the phases of such visibilities by calibrating those antennas, leaving all 
the other visibilities with the phases dispersed between $-\pi$ and $\pi$. 
In other words, the phases of the visibilities with large closure phases 
contribute to increase the value of the $\chi^2$ at the minimum, but the 
position of such minimum only depends on 
the visibilities with phases that can be modelled with a point source. That 
is, all the visibilities susceptible of producing a compact source are 
calibrated and their phases concentrate around zero. All the other 
visibilities tend to have their phases uniformly distributed between 
$-\pi$ and $\pi$, thus generating, after the Fourier inversion, a null mean 
flux density in the map. Thus, selfcal always produces a positive bias in the 
mean flux density of the map, given that selfcal only acts, effectively, 
on the phases that can be approached to zero, because their corresponding 
closure phases are close to zero.

One might think that it is very difficult for an antenna to be involved in a 
large number of closure phases close to zero, given that the distribution 
of phases is uniform. One might think that, in average, a given antenna 
is involved in the same number of closure phases that are close to zero than 
in the closure phases that are distant from zero. In such case, it would be 
impossible for selfcal to select which antenna should be calibrated, given 
that all the antennas participating in each scan would have the same chances 
for being calibrated. But this is only true in average. In the distributions 
of all the interferometric observations, there are statistical fluctuations, 
which are {\em always} used by selfcal for the generation of a spurious point 
source. For a given antenna $i$, the probability of finding $n$ closure phases 
(in which that antenna is involved) close to zero is:

\begin{equation}
P_{i}(n) = \frac{1}{2^{N'}}\frac{(N')!}{n! (N' - n)!}
\end{equation}

\noindent where $N' = (N-1)(N-2)/2$ is the total number of closure phases in 
which antenna $i$ is involed. (Notice that even though all closure phases 
are not independent, the closure phases with one common antenna are.) Thus, 
there is a finite probability of finding an 
antenna that appears in more than $N'/2$ closure phases close to zero. In 
such cases, the phases of the baselines in which the antenna $i$ appears will 
be minimized with success, generating a positive mean flux density in the map. 
Actually, even the cases in which $n < N'/2$ can also be used by selfcal for 
the generation of a spurious point source. In such cases, the closure phases 
in which antenna $i$ appears will tend to gather around $-\pi$ and $\pi$, 
meaning that there are other antennas that will appear in a large number 
of closure phases close to zero (i.e., the antennas belonging to the closure 
phases distant from zero in which antenna $i$ appears).

In short, there are always statistical fluctuations in the pure noise 
distribution of phases that can be used by selfcal to produce (by a selection 
process of the antennas most commonly being involved in the closure phases 
close to zero) a point source with a spurious source flux density.

It must be said that Global Fringe Fitting, when applied to noisy data, can 
also generate a spurious source from pure noise, given that this algorithm 
also finds antenna-based calibrations for adjusting the interferometric 
fringes of all the baselines in each scan. The spurious source is
generated as long as the minimum SNR of the fringes to be considered in the 
fit is set to a small value (lower than 2 or 3)\footnote{Even though a 
SNR of 2$-$3 is not generally used in the calibration of typical observations 
(being SNR cutoffs of $\sim$5 more common), low SNR cutoffs may be applied 
with very small delay/rate windows, say, after a phase-reference 
pre-calibration.}. In those cases, the Fringe 
Fitting would work on correlation peaks (fringes) produced, in
many cases, by spurious noise fluctuations. Then, by the same reasons given 
above, there would be a relatively high probability of generating a spurious 
point source from pure noise. 

\section{Dependence of the flux density of the spurious source on the 
characteristics of the observations}

In this section we consider how the flux density of the spurious source 
generated by selfcal depends on the parameters defining a set of 
observations. The parameters that we consider are the sensitivity of 
the array (for simplicity, we assume the same sensitivity for all the 
baselines of the interferometric array), the number of antennas of the 
interferometer, and the averaging time of the selfcal solutions. For the 
case of the Global Fringe Fitting algorithm, the averaging time of the 
solutions is equal to the duration of the scans.

\subsection{Numerical study}

We generated a set of synthetic interferometric data with the program 
{\sc fake} of the Caltech Package (Pearson \cite{Pearson1991}). We 
generated 6 hours of observations using a set of 20 antennas. All these 
antennas had the same diameters (25\,m) and the same system temperatures 
(60\,K). The correlator integration time was set to 2\,seconds. The source 
model used by {\sc fake} consisted on a single point source with a flux 
density of 1\,nJy (of course, completely undetectable by the 
interferometer). The data generated this way thus contain 6\,hours of pure 
noise observations made at 20 identical antennas under identical 
conditions. The mean of all the visibility amplitudes is 106\,mJy.

We used the program {\sc difmap} (Shepherd et al. \cite{Shepherd1995}) for 
hybrid mapping. We applied the natural weighting scheme to the visibilities, 
for sensitivity optimization, and applied an initial selfcal using a centered 
point source. The hybrid mapping steps were repeated until the $\chi^{2}$ of 
the fit of selfcal arrived to convergence. Then, we deconvolved the resulting 
point source using the CLEAN algorithm to see how much flux density was 
generated by selfcal. This process was repeated for different numbers of 
antennas and for different averaging times of the selfcal solutions. The 
spurious point source flux densities obtained in all these cases are shown 
in Fig. \ref{Resultados} as filled circles.

\subsection{Analytical study}

When the interferometer observes only noise, the sensitivity of the 
baselines defines the value of the standard deviation, $\sigma$, of the 
Gaussian distributions of the real and imaginary parts of the visibilities. 
As the sensitivity increases, the thermal noise of the baselines decreases, 
decreasing also the value of $\sigma$. Given that selfcal has to do only 
with phases and leaves the amplitudes unaltered, the recoverable flux density 
of the spurious source depends linearly on the rms of the amplitudes, 
which in turn also depends linearly on the value of $\sigma$. Thus, an 
increase in the sensitivity of the interferometer decreases the amount of 
spurious flux density recoverable from the data using selfcal. The constant 
factor in the ratio between the flux density recovered and the rms of the 
visibility amplitudes depends on the method used to estimate the flux density 
(i.~e. different deconvolution algorithms or modelfitting to the visibilities).
In our case, from the numerical study described in the previous section, we 
determine it to be $0.907 \pm 0.002$.

The averaging time of the selfcal solutions also has an effect on the 
spurious source flux density. If we have one observation every $t_{0}$ 
seconds (usually, $t_{0}$ is 2\,seconds) and find only one solution of 
selfcal every $t$ seconds, with $t > t_{0}$, then the spurious source flux 
density decreases. Finding a single solution of selfcal every $t$ seconds is
equivalent to averaging all the observations in bins of $t$ seconds and, 
afterwards, self-calibrating the resulting visibilities. When we average the 
visibilities in blocks of $t$ seconds, we are averaging separately the real 
and imaginary parts of the visibilities, which follow Gaussian distributions. 
The effect of this average is that the standard deviations of the resulting 
distributions decrease by a factor $\sqrt{t/t_{0}}$, because of the Central 
Limit Theorem.

The dependence of the spurious source flux density on the number 
of antennas is more difficult to find. There are lots of possible 
combinations of phases and baselines that help selfcal to generate a 
spurious source, and each one of these combinations has a different 
weight in the final spurious flux density. We can use a simplified 
phenomenological model to find out the recovered flux density as a function 
of the number of antennas. In principle, the recovered flux density 
depends directly on how well the point model fits the data. A good indicator 
of the {\em adjustability} of the data, with the phases randomly distributed, 
is, for each scan, equal to the number of visibility phases divided by the 
number of phase gains to fit. That is, the number of phases per parameter. 
When the number of phases per parameter increases, one single parameter must 
account for the minimization of more phases, and the quality of the fit 
decreases. In our case, the number of parameters is equal to $N - 1$, given 
that one antenna (the reference antenna) has a null phase gain by definition. 
Thus, the number of phases per gain to be fitted is:

\begin{equation}
\frac{\textrm{\#} phases}{\textrm{\#} gains} = \frac{N (N - 1)}{2} 
\frac{1}{N - 1} = \frac{N}{2}
\end{equation}

We can now look for a model of the dependence of the recovered spurious 
source flux density on the number of antennas using the quantity $N/2$ as 
variable. From the numerical simulations described in the previous section, 
we have found that the spurious source flux density is fitted very 
well using the simple model $F_{\textrm{\small sp}} \propto (N/2)^{\gamma}$, 
with $\gamma = -0.413 \pm 0.001$.

\begin{figure}[t]
\centering
\includegraphics[height=5.25cm,angle=0]{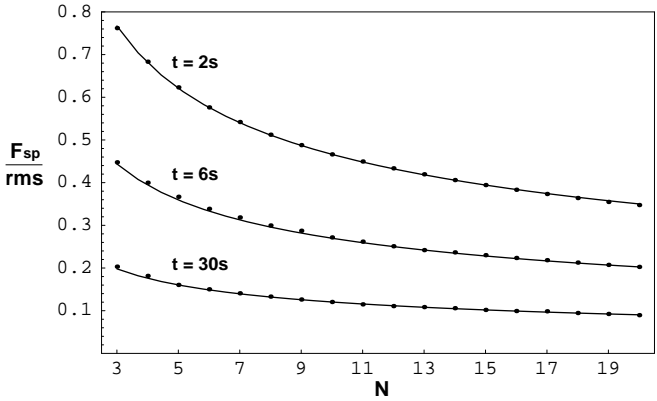}
\caption{Flux densities of an spurious point source, $F_{\textrm{\small sp}}$, 
in units of amplitude rms, $\rho$, recovered from hybrid 
mapping using pure noise synthetic visibilities, as a function of the number 
of antennas ($N$) and the averaging time of the selfcal solutions ($t$). 
The lines correspond to our model (Eq. \ref{ModeloFake}) and the 
dots to the numerical simulations.}
\label{Resultados}
\end{figure}

Taking these considerations into account, the recoverable flux density 
using selfcal on a set of randomly distributed data can be written as:

\begin{equation}
F_{\textrm{\small sp}} = 0.907 \rho \sqrt{\frac{t_0}{t}} \left (\frac{N}{2} 
\right )^{-0.413}
\label{ModeloFake}
\end{equation}

\noindent where $F_{\textrm{\small sp}}$ is the spurious source flux density 
that can 
be generated by selfcal, $\rho$ is the root-mean-square (rms) of the visibility 
amplitudes, $t_0$ is the averaging time used in the correlator (typically, 
$t_0$ = 2\,seconds), $t$ is the averaging time of the selfcal solutions, 
and $N$ is the number of antennas of the interferometer. We note that the 
duration of the whole set of observations does not affect 
$F_{\textrm{\small sp}}$, 
since this flux density depends on the ratio between the number of phases 
close to zero and the number of phases distant from zero, but does not depend 
on the total amount of visibilities used in the Fourier inversion. 
This model is shown in Fig. \ref{Resultados}.

\subsection{The use of selfcal in specific situations}

Equation \ref{ModeloFake} gives an estimate of the contribution of the 
artifacts of selfcal to the flux density of a source obtained by 
calibrating the antennas with the hybrid mapping technique. For cases of 
high SNR data, such contribution to the total flux density of the sources 
is negligibly small. However, when the flux density of a source is 
comparable to the rms of the visibility amplitudes, care must be exercised 
with the use of selfcal or the Global Fringe Fitting algorithm.

We note that in the worst situation for the use of selfcal (i.e., 3 
antennas and $t = t_0$) the amount of spurious source flux density is as 
large as 76\% of the rms of the visibility amplitudes. For 10 antennas (the 
case of the VLBA) the recoverable flux density decreases to 46\% of the rms 
(and can be lower if we set $t > t_0$). For interferometers with a large 
number of antennas, the amount of spurious source flux density is, of 
course, smaller. For example, if we extrapolate the results shown in Fig. 
\ref{Resultados} to 50 antennas (the case of ALMA), the spurious source flux 
density generated by selfcal would be 24\% of the rms of the visibility 
amplitudes, using $t=t_0$.

All these results assume the same sensitivity for all the baselines. In real 
cases, each baseline has its own sensitivity, with the longest baselines 
noisier than the shortest ones. The use of data from all the baselines in 
the fit can worsen the situation. A good alternative for avoiding the 
spurious source generated by selfcal or, at least, to make its flux density 
smaller is to flag or downweight the longest baselines in the fit and/or to 
increase the statistical weight of the data coming from the most sensitive 
antennas of the array. 
Nevertheless, even doing so, the statistical fluctuations of the closure 
phases will always tend to make, after the use of selfcal, a spurious source 
with a considerably large flux density. 

A better way to calibrate faint source data is using the 
phase-reference technique (e.g., Beasley \& Conway \cite{Beasley1995}). When 
using this technique, scans of a strong (reference) source are introduced 
between the scans of the faint (target) source. Then, the antenna gains are 
determined from the observations of the strong source and then interpolated 
to the scans of the faint (target) source. This technique is rather 
insensitive to the artifacts of selfcal and the probability of generating a 
fake signal from noise is practically zero. This is so, because the 
calibration of the target source comes from the analysis of data coming from 
another source (the reference source). Therefore, the noise in the data of 
the faint source does not affect the antenna calibrations. However, it is 
common to use the antenna gains determined from the phase reference as an 
a priori calibration, performing then a Global Fringe Fitting on the target 
source data using small search windows (based on the calibration from the 
reference source data) or applying self-calibration to the target 
visibilities in order to improve the dynamic range of the final image. 
In some cases, this might be malpractice, because the probability of 
generating a spurious source flux density from noise appears anew with full 
strengh, wasting all the benefits of the phase referencing.

\section{Tests of the reliability of a source detection from noisy data}

In this section we present two simple tests that can be performed on 
real data in order to check the reliability of a possible source detection, 
or to check if part of the flux density of a detected source may come from 
artifacts of selfcal. These tests are only meaningful when they 
are applied to extremely noisy data. However, the application of these 
tests to high SNR data can still give us an idea of the possible 
contribution of the artifacts of selfcal to the flux density recovered 
from a source. 

We assume, in all our discussion, that the detected source is compact 
enough to be considered point-like, without any important loss of precision. 
As in the previous sections, to simplify the expressions we also assume 
that all the baselines of the interferometer have the same sensitivity.

\subsection{Test based on the averaging time of the selfcal solutions}

The dependence of the spurious source flux density on the averaging time 
of the selfcal solutions is determined by the dependence of the standard 
deviation $\sigma$ on the averaging time. That dependence translates into 
equation \ref{ModeloFake}. However, the flux density of a real source in the 
data is independent of the averaging time of the selfcal solutions. 
We can use this condition to estimate the flux density of a (possibly real) 
source detected under critical circumstances. If we apply phase 
self-calibrations to a real data set for different averaging times, $t$, of 
the selfcal solutions, the flux densities recovered after each 
self-calibration, $F_{\textrm{\small self}}$, are given by the formula:

\begin{equation}
F_{\textrm{\small self}} = F_{\textrm{\small sp}} + F_{\textrm{\small real}} = 
\frac{K}{\sqrt{t}} + F_{\textrm{\small real}}
\label{SelfcalTime}
\end{equation}

\noindent where $K$ (related to the specifics of the interferometer) and 
$F_{\textrm{\small real}}$ (an estimate of the flux density of the real source) 
are parameters to be fitted.
 
\subsection{Test based on the distribution of closure phases}

In the case that the signal of a real source is included in the 
visibilities, the probability distributions of the real and imaginary 
parts are still Gaussian, but the mean value of the real part of the 
visibilities (if such visibilities are well calibrated) will be equal to 
the flux density of the real source. Then, it can be easily shown that the 
probability distribution of the resulting visibility amplitudes and phases 
is:

\begin{equation}
g(A,\phi) = \frac{A}{\sigma^2} 
\exp{\left (-\frac{(A+F)^2}{2\sigma^2} \right )} 
\exp{\left (\frac{A F \cos\phi}{\sigma^2} \right )}
\label{ProbDistri2}
\end{equation}

\noindent where $g(A,\phi)$ is the probability density of amplitudes 
(variable $A$) and phases (variable $\phi$), $F$ is the flux density of the 
real source, and $\sigma$ is the width of the distributions of the real and 
imaginary parts of the visibilities.  

Equation \ref{ProbDistri2} turns into equation \ref{ProbDistri1} for $F = 0$. 
When $F$ is different from zero, the distribution of phases is not uniform. 
As $F$ increases, the phases gather around zero in a Gaussian-like manner. 
The distribution of amplitudes also changes, increasing the ratio 
between the rms of the visibility amplitudes and $\sigma$.

How could the information provided by equation \ref{ProbDistri2} be used to 
check the reliability of a possible source detection? The closure phases are 
robust quantities that can be used to check the reliability of a source 
detection. The closure phases are sensitive to $F$ and are not affected 
by the phase self-calibration. The distribution of closure phases tends 
to gather around zero if there is signal of a real source in the data (and 
particularly so if the source has no structure) and is uniformly 
distributed if there is only noise in the data. From the definition of 
closure phase (see equation \ref{ClosPhase}) we conclude that, if the 
visibility phases are well calibrated, the probability distribution of 
closure phases is equal to: 

\begin{equation}
c(\beta) = \int^{\pi}_{-\pi}{\int^{\pi}_{-\pi}{p(\phi_{1}) p(\phi_{2}) 
p(\beta - \phi_{1} - \phi_{2}) d\phi_{1} d\phi_{2}}}
\label{ClosDistri}
\end{equation}

\noindent where $c(\beta)$ is the probability density of the closure phase, 
$\beta$, and

\begin{equation}
p(\phi) = \int^{\infty}_{0}{g(A,\phi) dA}
\end{equation}

Let us ellaborate on this: for the case of a perfect calibration, the phases 
of data from a given baseline $(i,j)$ are independent of the phases 
from any other baseline, given that all the contributions to $\phi_{ij}$ 
come from noise, which is uncorrelated between the different 
baselines. 
Thus, the probability distribution function of any linear combination of 
visibility phases (as it is the case of the closure phase) is equal 
to the product of the probability distributions of the visibility phases. 
However, for the case of a non-perfect calibration, in which selfcal has 
introduced a spurious source flux density in the data, the distributions 
of the phases from the different baselines will no longer be independent 
and equation \ref{ClosDistri} will not quite apply. In such cases, the phase 
distributions, $p(\phi)$, will be more peaked around zero, but correlations 
will appear among the phases of the different baselines. Therefore, the 
probability distribution function of a linear combination of phases will 
not be equal to a simple product of $p(\phi)$. Nonetheless, the 
correlations between baselines that selfcal introduces in the 
data keep the distribution of closure phases, $c(\beta)$, unaltered. 
That is, even after a selfcal iteration has generated a spurious source in 
the map, the distribution of closure phases is still equal to $c(\beta)$, as 
computed from the distribution of the phases, $p(\phi)$, corresponding to 
perfectly calibrated data.

\begin{figure*}[t]
\centering
\includegraphics[height=5.25cm,angle=0]{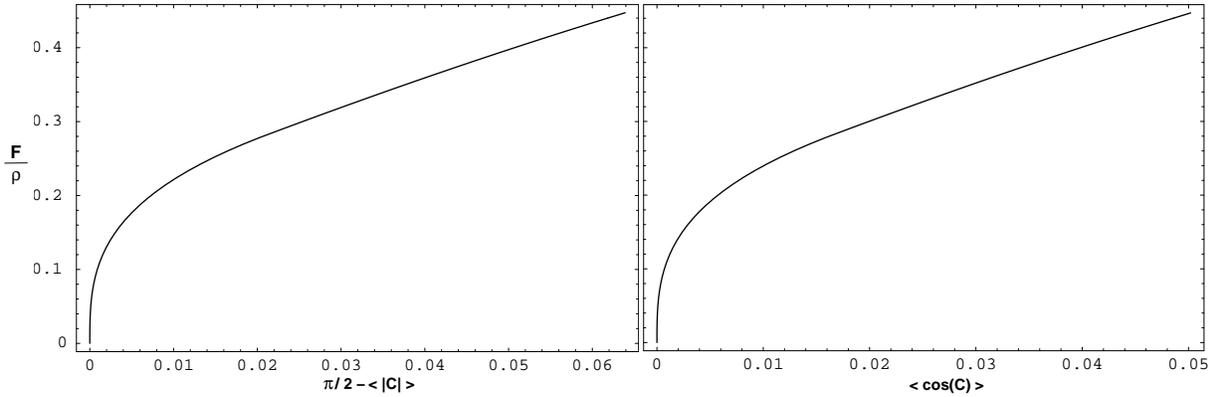}
\caption{Theoretical dependence of the flux density of a real observed source 
(in units of the rms of the visibility amplitudes), as a function of ``$\pi/2$ 
minus the mean absolute values of the closure phases'' (left side) and as a 
function of the mean cosine of the closure phases (right side). These 
theoretical behaviors have been obtained from equation \ref{ClosDistri}.}
\label{RelaTeo}
\end{figure*}

This invariance property of the closure phase distribution, $c(\beta)$, can 
be used to check the reliability of a possible source detection.  
In Fig. \ref{RelaTeo}, the theoretical flux density of a real source (in 
units of the rms of the amplitudes, $\rho$) is shown as a function of the 
mean absolute value of the closure phases and as a function of the mean 
cosine of the closure phases. Both quantities are directly related to the 
deviation of the closure phase distribution from a uniform distribution. 
If the closure phases are uniformly distributed between $-\pi$ and $\pi$, 
the average of their absolute values will be equal to $\pi/2$ and the mean 
cosine will be equal to zero. Any concentration of closure phases around 
zero (i.e., the existence of any real source in the data) will result 
in a decrease of the first value and an increase of the second.

If a faint source is detected and the reliability of the detection needs to be 
checked, the average of the absolute values of the closure phases of the 
observations (or the average of their cosines) can be computed and compared 
in Fig. \ref{RelaTeo} to determine how concentrated around zero the 
closure phases are. This way, it can be seen whether the flux density 
of our detection is consistent with a real source or not.

\subsubsection{Comments on the closure-phases test}

At this stage, it must be said that, even though the self-calibration does 
not affect the closure phases, a {\em time average} of a set of 
self-calibrated visibilities {\em will} change the closure phase 
distribution. That is, if selfcal generates a spurious source from noise 
and the visibilities are averaged in time bins of $t$ seconds, with $t>t_0$, 
the closure phase distribution changes and the resulting closure phases 
concentrate around zero, creating the effect of a source that is
completely indistinguishable from a real one using any test.

For very small values of the flux density $F$, this test is not as good 
as the first one. As we can see in Fig. \ref{RelaTeo}, the closure 
phases do not dramatically change their distribution for flux densities in 
the range between 0 and $\sim$15\% of the rms of the visibility amplitudes. 
For tentative detections under critical circumstances in that range, this 
test could lead us to the wrong conclusion about the reliability of a source 
detection. Looking at it from a different viewpoint, Fig. \ref{RelaTeo} 
provides an interesting lesson: {\em for a given dataset, there can exist a 
real faint source (appearing in a map with a dynamic range of 6 or more) 
even if the distribution of closure phases is uniform, that is, even if the 
closure phase distribution is noise-like}. Thus, a conclusion on the 
reliability of a source detection based only on the closure phase 
distribution being extremely noisy, is not definitive.

Another thing worth-noticing is that this test assumes the same 
sensitivity for all the antennas and a source compact enough for 
generating closure phases close to zero even for the longest baselines (which 
may not be the case, specially for sources with a very low flux density per 
unit beam). These assumptions impose limitations to the use of this 
test. However, it could still be applied by restricting its use to the 
shortest baselines with similar antennas. For an array with a large antenna, 
the closure phases in which that antenna appears could still be used (for 
sensitivity optimization in the closure phase distribution), but then the flux 
density estimated from the amplitude rms might have a bias produced by the 
very different rms in different baselines.

We must also note that in cases of high SNR, the rms of the visibility 
amplitudes is no longer related only to the thermal noise of the 
baselines (the flux density of the source affects the value of the rms), 
and the fraction $F/\rho$ shown in Fig. \ref{RelaTeo} should be accordingly 
corrected. For cases of high SNR, the quantity that will substitute $\rho$ in 
the fraction $F/\rho$ shown in Fig. \ref{RelaTeo} is $\sqrt{\rho^2-F^2}$.

\subsection{Application to real cases}

Real data do not obey the simplifying assumptions that we have used in the 
earlier sections. The baselines of a real interferometer have different 
sensitivities, which also vary in time. Thus, in order to check the 
reliability of a source detection from real data we must search a subset 
of observations in which the sensitivity of the antennas is approximately 
constant. Moreover, we must only work with the subset of most sensitive 
antennas of the interferometer. If there is one antenna in our interferometer 
that is clearly more sensitive than the others, we should compute only 
the average of the closure phases in which this more sensitive antenna 
appears, in order to insure the possible signature of the source in the 
closure phase distribution. In what follows, we apply our reliability 
criteria to real data corresponding to the radio supernova SN\,2004et 
(Mart\'i-Vidal et al. \cite{MartiVidal2007}).

\subsubsection{Supernova SN\,2004et}

We observed this supernova on 20 February 2005. From all data reported in
Mart\'i-Vidal et al. (\cite{MartiVidal2007}), we have chosen the following 
subset of antennas: Brewster, Fort Davis, Green Bank, Hancock, Kitt Peak, 
and Owens Valley. We have only computed the closure phases in which the 
antenna Green Bank appears, and we have used data only from 14\,hr to 
20\,hr (UT). These choices are based on the quality of the data for our 
purposes (i.e., the stability of the antenna sensitivities, which we assume 
proportional to the system temperatures registered for each station).

First test: we self-calibrated the SN\,2004et data using different averaging 
times, ranging from 2 to 120\,seconds (roughly, the duration of one scan).
The fit of the flux densities recovered from the SN\,2004et data as a function 
of the averaging time of the selfcal solutions, equation \ref{SelfcalTime}, 
results in a value of $F_{real} = 0.90\pm0.13$\,mJy. This value is 
clearly higher than zero, indicating that there is a real signal in the data. 
This value is also close to the flux density of SN\,2004et reported by 
Mart\'i-Vidal et al., $0.87\pm0.03$\,mJy, recovered from phase referenced data 
with a deconvolution using CLEAN.

Second test: even though we know that it is not quite appropriate, we have also 
performed this test on the SN\,2004et data. The flux density of SN\,2004et is 
too low compared to the rms of the visibility amplitudes ($\sim$20\,mJy) for 
obtaining a good result with the test of the closure phase distribution.
The average value of the cosines of all the closure phases considered in 
these observations is equal to $0.005\pm0.002$. This average value is 
slightly higher than 0. The average of the absolute values of the closure 
phases is $0.0065\pm0.0020$.
From Fig. \ref{RelaTeo} we estimate a flux density of the supernova of 
$0.18\pm0.08$ times the rms of the visibility amplitudes, $\rho$, used in 
our computations, which, as said above, is $\sim20$\,mJy. Hence, the estimated 
flux density of the (real) source in the data is, then, $3.8\pm1.6$\,mJy. This
value is too high, but compatible (at a 2--sigma level) with the flux 
density estimated from the other reliability test. As expected, also the 
uncertainty of the flux density estimated is much higher in this case.

Thus, the reliability tests for this source are successful. We must note 
that Mart\'i-Vidal et al. calibrated the data of SN\,2004et using the 
phase-reference technique, in which they interpolated the antenna gains 
obtained from the observations of the source J2022+614 to the scans of the 
supernova. These authors did not refine afterwards such a calibration 
applying a Global Fringe Fitting to the supernova data. This procedure 
assured a reliable detection of the supernova. These authors did not apply 
any other calibration (selfcal) to the phased-referenced supernova data, 
in order to avoid any possible artifact introduced by the use of selfcal.

\section{Conclusions}

We have analyzed the consequence of the phase-self-calibration algorithm when
it is applied to extremely noisy data. We have studied how this algorithm 
and the statistical fluctuations of the visibility phases can create a 
spurious source from pure noise. The flux density of the spurious source can 
be a considerable fraction of the rms of the visibility amplitudes. 
The application of other other antenna-based calibration algorithms (like the 
Global Fringe Fitting) to noisy data can have similar consequences to 
those of selfcal if the SNR cutoff of the gain solutions is set to small 
values.

We have considered numerical and analytic studies to show how the 
flux density of a spurious source created by selfcal depends on the 
number of antennas, the sensitivity of the array, and the averaging time 
of the selfcal solutions. We have also presented two simple tests that can 
be applied to real data in order to check if the detection of a faint 
source could be the result of the application of an antenna-based calibration 
algorithm to noisy data. These tests basically relate the averaging time 
of the selfcal solutions and the characteristics of the closure phase 
distribution to the flux density of a compact source possibly present in the 
data. To show a practical case, we have applied these tests to a set of real 
VLBI observations of supernova SN\,2004et and found good agreement between 
the flux density recovered by CLEAN from the (phase-referenced) visibilities 
of this supernova (Mart\'i-Vidal et al. \cite{MartiVidal2007}) and the flux 
density estimate provided by our reliability tests.

\begin{acknowledgements}
This work has been partially funded by Grants AYA2004-22045-E and 
AYA2005-08561-C03-03 of the Spanish DGICYT. We agree the anonymous referee
for his/her very helpful comments, corrections, and suggestions.
\end{acknowledgements}

\end{document}